\documentclass[12pt]{article}
\usepackage{graphicx} 
\usepackage{hyperref}
\usepackage{url}
\usepackage{listings,jvlisting}
\usepackage{amssymb}
\usepackage{tabularx}
\usepackage{amsmath}
\usepackage{mdframed}
\usepackage{graphicx}
\usepackage{setspace}
\usepackage{here}
\usepackage{authblk}
\usepackage{mathtools}
\usepackage{amsthm} 
\usepackage{algorithmic}
\usepackage{algorithm}
\usepackage{cite}
\usepackage{comment}
\usepackage{colortbl,array,xcolor}

\newcommand{\indep}{\mathop{\perp\!\!\!\perp}}

\newcommand{\bld}{\boldsymbol}

\def\be{{\beta}}
\def\ga{{\gamma}}

\def\ep{{\varepsilon}}

\def\si{{\sigma}}

\def\bbe{{\text{\boldmath $\beta$}}}
\def\bga{{\text{\boldmath $\gamma$}}}

\def\Si{{\Sigma}}

\def\x{{\text{\boldmath $x$}}}

\def\z{{\text{\boldmath $z$}}}

\def\ph{{\hat p}}

\def\Wh{{\hat W}}

\def\Eh{{\widehat E}}

\def\Nc{{\cal N}}

\begin{document}
\title{Sample size re-estimation in blinded hybrid-control design using inverse probability weighting}
\author[1]{Masahiro Kojima\footnote{Address:1-13-27 Kasuga,Bunkyo-ku,Tokyo 112-8551, Japan. Tel: +81-(0)3-3817-1949 \quad
E-Mail: mkojima263@g.chuo-u.ac.jp}}
\author[2]{Shunichiro Orihara}
\author[3]{Keisuke Hanada}
\author[4]{Tomohiro Ohigashi}
\affil[1]{Department of Data Science for Business Innovation, Chuo University}
\affil[2]{Department of Health Data Science, Tokyo Medical University}
\affil[3]{Faculty of Medicine, Wakayama Medical University}
\affil[4]{Department of Information and Computer Technology, Faculty of Engineering, Tokyo University of Science}

\maketitle
\abstract{\noindent
With the increasing availability of data from historical studies and real-world data sources, hybrid control designs that incorporate external data into the evaluation of current studies are being increasingly adopted. In these designs, it is necessary to pre-specify during the planning phase the extent to which information will be borrowed from historical control data. However, if substantial differences in baseline covariate distributions between the current and historical studies are identified at the final analysis, the amount of effective borrowing may be limited, potentially resulting in lower actual power than originally targeted. In this paper, we propose two sample size re-estimation strategies that can be applied during the course of the blinded current study. Both strategies utilize inverse probability weighting (IPW) based on the probability of assignment to either the current or historical study. When large discrepancies in baseline covariates are detected, the proposed strategies adjust the sample size upward to prevent a loss of statistical power. The performance of the proposed strategies is evaluated through simulation studies, and their practical implementation is demonstrated using a case study based on two actual randomized clinical studies.
}
\par\vspace{4mm}

\section{Introduction}

The utilization of external data has been advancing, as exemplified by the issuance of guidance from the FDA regarding the use of external controls~\cite{FDA2023ExternallyControlled}. While randomized controlled studies are well established as the gold standard for evaluating efficacy, there has been growing interest in hybrid control designs, which incorporate external data into the control arm of randomized controlled studies. These approaches have stimulated considerable discussion around methods for borrowing information from external sources~\cite{ventz2022design,sengupta2023emulating,guo2024adaptive,ohigashiUsingHorseshoePrior2022,ohigashiNonparametricBayesianApproach2024,ohigashiPotentialBiasModels2025}.

As a sample size re-estimation (SSR) method for clinical studies employing hybrid control designs, \cite{mutze2018sample} proposed an approach that utilizes a meta-analytic predictive prior for the outcome variance, allowing sample size re-estimation at interim analysis based on observed variability. \cite{brakenhoff2019bayesian} introduced a method that uses a power prior to downweight historical information when the pre-specified variance differs from the observed variance, thereby enabling re-estimation. \cite{lin2022dynamic} developed a method that performs matching at interim analysis and conducts SSR using conditional power, as is common in conventional SSR approaches. \cite{lu2024propensity} stratified patients using propensity scores and determined the degree of borrowing based on the number of matched patients within each stratum. However, none of these studies provides detailed guidance on the practical implementation of hybrid control designs. Furthermore, all of these methods require unblinding of treatment group information at the time of sample size re-estimation. Although \cite{hees2017blinded} proposed a blinded SSR method, it is limited to binary outcomes and still requires observation of the number of events.

In this paper, within the context of a double-blind, two-arm hybrid control design, we propose two sample size re-estimation strategies that utilize inverse probability weighting (IPW), where the weights are based on the inverse of the probability of being assigned to either the historical or current study. The first method is intended for use by unblinded statistician, such as members of an independent data monitoring committee, under conditions where only outcome data are unblinded for only unblinded statistician, while the treatment group assignment in the current study remains blinded for investigators, participants, and statisticians. However, in non-inferiority studies, outcomes may be unblinded. In this setting, sample size re-estimation is performed in a mechanistic manner based solely on outcome information. The second strategy maintains blinding of the outcomes and instead conducts sample size re-estimation using only baseline covariate information. By eliminating the need for an unblinded statistician, this strategy makes SSR easy to implement.

This paper is organized as follows. Section 2 defines the weighting scheme as a preliminary. Section 3 introduces the two proposed strategies for sample size re-estimation. Section 4 presents the simulation settings and results. Section 5 provides a case study demonstrating sample size re-estimation based on two actual studies. Finally, Section 6 discusses the findings.

\section{Preliminaries}
We consider a blinded, randomized, two-arm controlled study in which $A=1$ denotes the treatment group and $A=0$ denotes the control group. Additionally, we incorporate a hybrid design that includes historical control data, indicated by $R=0$. Here, $R$ denotes whether a subject belongs to the current study ($R=1$) or to the historical study ($R=0$). Therefore, the study comprises three groups: a current treatment group ($A=1$ and $R=1$), a current control group ($A=0$ and $R=1$), and a historical control group ($A=0$ and $R=0$).

Our goal is to estimate the average treatment effect (ATE) within the current study population, defined as
$$
{\rm E}\left[Y_{1}-Y_{0}\mid R=1\right]={\rm E}\left[Y_{1}\mid R=1\right]-{\rm E}\left[Y_{0}\mid R=1\right]=:\theta_{1}-\theta_{0},
$$
where $Y_{a}$ ($a = 0, 1$) denotes the potential outcome. We assume that the treatment assignment $A \in \{0,1\}$ is randomized within the current study: $A \indep (Y_{1},Y_{0}) \mid R=1$, an assumption known as exchangeability \cite{He2020}. Under this assumption, the ATE can be identified from observed data:
$$
{\rm E}\left[Y \mid A=1, R=1\right] - {\rm E}\left[Y \mid A=0, R=1\right] = {\rm E}\left[Y_{1} \mid R=1\right] - {\rm E}\left[Y_{0} \mid R=1\right],
$$
where $Y := A Y_{1} + (1 - A) Y_{0}$ for subjects in the current study.

To improve the precision of the treatment effect estimate, we seek to incorporate information from the historical control group. However, care must be taken when selecting historical control, as their baseline characteristics may differ from those of subjects in the current study. In the following sections, we propose two strategies to address this issue while maintaining the blinding of the study.

To formalize the framework, we introduce a set of baseline covariates, denoted by $X \in \mathbb{R}^{p}$. Let $n_{c}$ and $n_{h}$ represent the sample sizes of the current and historical data, respectively, with $n = n_{c} + n_{h}$. Additionally, let $i$ denote the subject ID, and assume that subjects are independently sampled from the super-population corresponding to each dataset.

\subsection{Data fusion of current and historical studies}
To combine the current and historical control group data, we assume a conditional independence assumption: $R \indep Y_{0} \mid X$, which notation was introduced by \cite{dawid1979conditional}. This assumption is analogous to exchangeability, in that the allocation mechanism $R$ is fully explained by the observed covariates $X$. Under this assumption,
\begin{align*}
{\rm E}\left[\frac{(1-R){\rm Pr}(R=1\mid X)Y_{0}}{{\rm Pr}(R=0\mid X)}\right]&=\sum_{r=0}^{1}\int\frac{(1-r){\rm Pr}(R=1\mid x)y_{0}}{{\rm Pr}(R=0\mid x)}f(y_{0},x,r)dy_{0}dx\\
&=\int\frac{{\rm Pr}(R=1\mid x)y_{0}}{{\rm Pr}(R=0\mid x)}f(y_{0}\mid R=0,x){\rm Pr}(R=0\mid x)f(x)dy_{0}dx\\
&=\int{\rm Pr}(R=1\mid x)y_{0}f(y_{0}\mid R=1,x)f(x)dy_{0}dx\\
&={\rm Pr}(R=1){\rm E}\left[Y_{0}\mid R=1\right].
\end{align*}


Therefore, using the following weight:
$$
W^{r=0} := 0.5 \times \frac{1}{{\rm Pr}(R = 1)} \cdot \frac{(1 - R) \, {\rm Pr}(R = 1 \mid X)}{{\rm Pr}(R = 0 \mid X)},
$$
the differences between the current and historical studies can be adjusted as ${\rm E}\left[Y_{0}\mid R=1\right]={\rm E}\left[W^{r=0} Y_{0}\right]$. Similarly, define the weight:
$$
W^{r=1} := 0.5 \times \frac{R}{{\rm Pr}(R = 1)}.
$$
Here, the factor $0.5 \times$ represents the “fair” use of information; it implies that the current and historical datasets are combined in a balanced manner, without prioritizing either dataset based on a specific objective. By incorporating historical data ($R = 0$), we expect an increase in the efficiency of ATE estimation:
\begin{align*}
\hat{\rm E}[Y_{0} \mid R = 1]
&= \frac{1}{n} \sum_{i = 1}^{n} \frac{0.5}{\widehat{\rm Pr}(R = 1)} \left[ \frac{R_{i}(1 - A_{i})}{\widehat{\rm Pr}(A = 0 \mid R = 1)} + \frac{(1 - R_{i}) \, \widehat{\rm Pr}(R = 1 \mid X_{i})}{\widehat{\rm Pr}(R = 0 \mid X_{i})} \right] Y_{i} \\
&= \frac{1}{n} \sum_{i = 1}^{n} \left[ \frac{(1 - A_{i})}{\widehat{\rm Pr}(A = 0 \mid R = 1)} \hat{W}_{i}^{r=1} + \hat{W}_{i}^{r=0} \right] Y_{i},
\end{align*}
where $\widehat{\rm Pr}(R = 1) = \frac{1}{n} \sum_{i = 1}^{n} R_{i}$, $\widehat{\rm Pr}(A = 0 \mid R = 1) = \frac{1}{n_{c}} \sum_{i = 1}^{n_{c}} (1 - A_{i})$, and $\widehat{\rm Pr}(R = 1 \mid X_{i})$ is estimated using methods such as logistic regression. Note that for the historical data, $Y \equiv Y_{0}$, as all subjects had no chance of being assigned to the study treatment.

\section{Proposed Sample Size Re--estimation methods and Test Statistics}
In this section, we propose two SSR strategies. The first is a conventional approach: using the outcome variance estimated from the data accrued up to the interim analysis, the sample size is recalculated. However, from a study integrity standpoint, relying on outcome information is generally discouraged. Therefore, the second SSR strategy does not utilize outcome data but instead relies solely on the variance of the weights. This latter approach is motivated by \cite{shook2022power}.

\subsection{SSR strategy 1:\ using outcome variance}
The first proposed strategy is grounded in the concept of sample size re-estimation (SSR) under blinded conditions\cite{Ki2003,Fr2006}. Specifically, the sample size is recalculated as
$$
\hat{N}_{1}=2(z_{1-\alpha}+z_{1-\beta})^2\frac{S_{1}^2}{\delta^2},
$$
where $\alpha$ and $\beta$ represent type I and I\hspace{-1.2pt}I error rates, respectively; $z_{\cdot}$ is the z-value (i.e., assuming z-statistics), and $\delta$ is the expected effect size initially identified before the study. Here, $S_{1}^2$ is estimated as a weighted one-sample variance:
\begin{align}
\label{S_est1}
S_{1}^2=\frac{\sum_{s=0}^{1}\sum_{j=1}^{n_{1i}}\hat{W}_{j}^{r=s}\left(Y_{j}-\bar{Y}\right)^2}{\left(\sum_{s=0}^{1}\sum_{j=1}^{n_{1s}}\hat{W}_{j}^{r=s}\right)-1},
\end{align}
where $n_{11}$ and $n_{10}$ are the numbers of subjects recruited at the time of re-estimation for the treatment and control group, respectively (i.e., $n_{1}=n_{11}+n_{10}<n$), and
$$
\bar{Y}=\frac{\sum_{s=0}^{1}\sum_{j=1}^{n_{1i}}\hat{W}_{j}^{r=s}Y_{j}}{\sum_{i=0}^{1}\sum_{j=1}^{n_{1i}}\hat{W}_{j}^{r=s}}.
$$
Note that various strategies exist for estimating $S_{1}^2$ are proposed, however, simple estimators like (\ref{S_est1}) perform well compared to more complex methods\cite{Fr2006}.

\subsection{SSR strategy 2:\ using only weight variance}
The second proposed strategy is a straightforward extension of the sample size estimation method \cite{shook2022power}. A key advantage of this approach is that it does not require any information about the outcome variable, which is expected to offer benefits in terms of both study integrity and operational simplicity. Specifically, the sample size is recalculated as
\begin{align}
\label{S_est2}
\hat{N}_{2} = (1 + k)(z_{1-\alpha} + z_{1-\beta})^2 \frac{\hat{\sigma}_{1}^2 / k + \hat{\sigma}_{0}^2}{\delta^2},
\end{align}
where $k = \widehat{\rm Pr}(R = 1) / \widehat{\rm Pr}(R = 0)$,
$$
\hat{\sigma}_{1}^2 = \sigma_{1}^2 \cdot \frac{ \widehat{\rm Pr}(R = 1) \cdot \frac{1}{n_{1}} \sum_{i = 1}^{n_{1}} R_{i} \left( \hat{W}_{i}^{r = 1} \right)^2 }{ \left( \frac{1}{n_{1}} \sum_{i = 1}^{n_{1}} R_{i} \hat{W}_{i}^{r = 1} \right)^2 }, \quad
\hat{\sigma}_{0}^2 = \sigma_{0}^2 \cdot \frac{ \widehat{\rm Pr}(R = 0) \cdot \frac{1}{n_{1}} \sum_{i = 1}^{n_{1}} (1 - R_{i}) \left( \hat{W}_{i}^{r = 0} \right)^2 }{ \left( \frac{1}{n_{1}} \sum_{i = 1}^{n_{1}} (1 - R_{i}) \hat{W}_{i}^{r = 0} \right)^2 },
$$
and $\sigma_{r}^2$ ($r = 0, 1$) denote the expected variances for the current and historical data, respectively, as specified at the design stage.

\subsection{Test statistics:\ random sampling from historical control}\label{sec:test}
In the final analysis, the following test statistic is used to evaluate the endpoint under the null hypothesis ($\theta_{1} - \theta_{0} = \tau_{0}$):
\begin{align}
\label{Test1}
\frac{\frac{1}{\hat{N}_{j}} \sum_{i=1}^{\hat{N}_{j}} \left[ \frac{A_{i}}{\widehat{\rm Pr}(A=1)} - \left\{ \frac{(1 - A_{i})}{\widehat{\rm Pr}(A = 0 \mid R = 1)} \hat{W}_{i}^{r=1} + \hat{W}_{i}^{r=0} \right\} \right] Y_{i} - \tau_{0}}{\sqrt{ \frac{\hat{\sigma}^{*2}}{\hat{N}_{j}} }} \quad j = 1, 2,
\end{align}
where
$$
\hat{\sigma}^{*2} = \frac{1}{\hat{N}_{j}} \sum_{i=1}^{\hat{N}_{j}} \left[ \frac{A_{i}}{\widehat{\rm Pr}(A = 1)^2} (Y_{i} - \hat{\theta}_{1})^2 + \left\{ \frac{(1 - A_{i})}{\widehat{\rm Pr}(A = 0 \mid R = 1)} \hat{W}_{i}^{r=1} + \hat{W}_{i}^{r=0} \right\}^2 (Y_{i} - \hat{\theta}_{0})^2 \right],
$$
$\hat{\theta}_{1} = \frac{1}{\hat{N}_{j}} \sum_{i=1}^{\hat{N}_{j}} \frac{A_{i}}{\widehat{\rm Pr}(A = 1)} Y_{i}$, and
$$
\hat{\theta}_{0} = \frac{1}{\hat{N}_{j}} \sum_{i=1}^{\hat{N}_{j}} \left[ \frac{(1 - A_{i})}{\widehat{\rm Pr}(A = 0 \mid R = 1)} \hat{W}_{i}^{r=1} + \hat{W}_{i}^{r=0} \right] Y_{i}.
$$
The test statistic in (\ref{Test1}) is derived from an estimating equation based on the M-estimator framework \cite{St2002}. For further details, see Appendix \ref{appa}.

The two proposed strategies update the sample size using similar forms, as shown in (\ref{S_est1}) and (\ref{S_est2}). However, the rationale behind each updating method differs substantially. To examine their properties, we present simulation studies and a real data application in the following sections.

\section{Simulation}
We conducted simulation studies to evaluate the performance of the proposed methods. In these simulations, we assessed the Type I error rate under the null hypothesis of no true treatment effect difference between the current control and treatment groups, and statistical power under scenarios where a true difference exists between the treatment and control group. The main assumptions for data generation were informed by the results of the ADC-016 and ADC-027 studies, which targeted patients with Alzheimer’s disease. In addition, we prepared alternative scenarios with slightly modified settings based on the reference information to assess the robustness of our proposed method. This dataset is used in the subsequent case study, with further details provided in Section 5. Summary statistics used as references from each study are presented in Section~\ref{sec:case}.
\subsection{Data-generating mechanisms}
The data-generating model used in the simulations is described below:
\begin{align}
y_{si} = \x_{si}^\top \bbe + \ep_{si}, \quad \ep_{si} \sim \Nc(0, \si_s^2),
\end{align}
where $s = c$ denotes the current study, $s= h$ denotes the historical study, and $i$ indexes individual subjects. The covariate vector is defined as $\x_{si} = (1, a_{si}, x_{si1}, x_{si2}, x_{si3}, x_{si4})^\top$, where the first element represents the intercept and is set to 1 for all subjects. The variable $a_{si}$ equals 1 if the $i$-th subject received the treatment and 0 otherwise. The covariates are generated as follows: $x_{si1} \sim \Nc(\mu_{1s}, \si^2_{1s})$, $x_{si2} \sim \mathrm{Ber}(p_s)$, $x_{si3} \sim \Nc(\mu_{2s}, \si^2_{2s})$, and $x_{si4} \sim \Nc(\mu_{3s}, \si^2_{3s})$. The true values of $\mu_{1s}$, $\mu_{2s}$, $\mu_{3s}$, and $p_s$ were specified separately for the historical and current studies to reflect distributional differences between the two populations. The parameters for the current study were fixed as follows: $\si^2_c = 10^2$, $\mu_{1c} = 75$, $p_c = 0.5$, $\mu_{2c} = 14$, $\mu_{3c} = 21$, $\si^2_{1c} = 8.5^2$, $\si^2_{2c} = 2.8^2$, and $\si^2_{3c} = 3.6^2$. Only the parameters for the historical study were varied across scenarios. Scenario 1 was designed under the assumption that the covariate distributions were identical between the current and historical studies. Scenario 2 used covariate distributions similar to those in the later case study. Scenarios 3 and 4 were constructed to investigate how the results of SSR would change when the covariate distributions differed more substantially. Scenario 5 assumed that both the variances of the covariates and the outcome differed greatly between the current and historical studies. The coefficient vector was set to $\bbe = (\be_0, \be_1, \be_2, \be_3, \be_4, \be_5)^\top$, where the intercept $\be_0$ was set to $1$. The treatment effect coefficient $\be_1$ was set to 0 for the Type I error rate evaluation and to $3.5$ for the power evaluation. The remaining coefficients were fixed at $\be_2 = 1$, $\be_3 = 1$, and $\be_4 = 1$.

The true propensity score model used for weighting was defined as:
\begin{align}
\Pr(r_i = 1 \mid \z_i) = \frac{1}{1 + \exp(-\z_i^{\top} \bga)},
\end{align}
where $r_i$ indicates whether the $i$-th subject belongs to the current study ($r_i = 1$) or the historical study ($r_i = 0$), $\z_i = (1, x_{si1}, x_{si2}, x_{si3}, x_{si4})^\top$, and $\bga = (\ga_0, \ga_1, \ga_2, \ga_3, \ga_4) = (2.030, -0.019, -0.106,$\\
$ 0.095, -0.009)$.

\begin{figure}[H]
    \centering
    \includegraphics[width=0.75\linewidth]{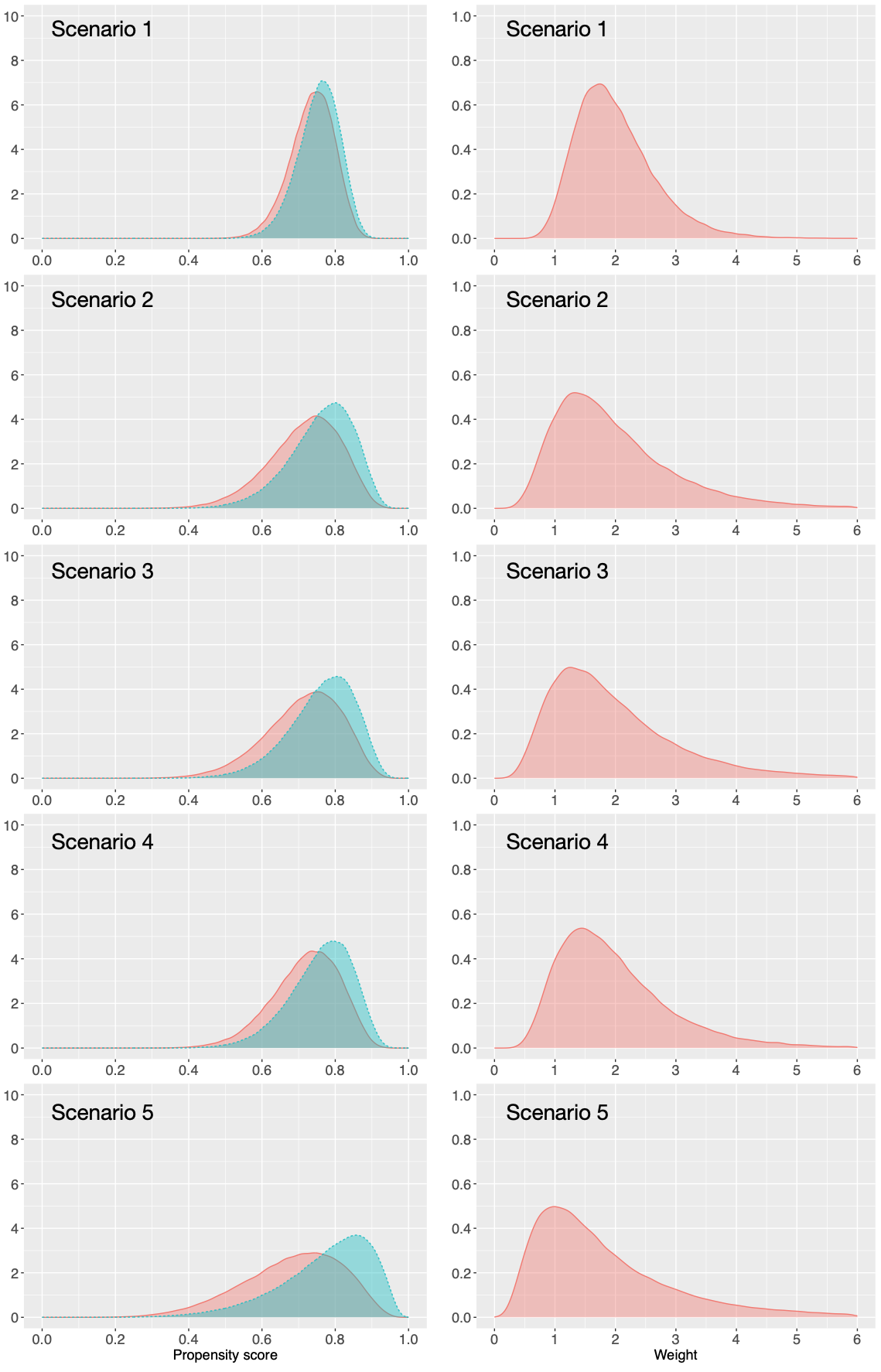}
    \caption{Distributions of propensity scores and weights for each scenario}
    \label{fig:Sim_Box_plot}
        {\footnotesize Green indicates the current trial (dashed line), and red indicates the historical trial. Regarding the weight, the current trial used fixed values, which were not illustrated in the figure. Specifically, the average weight in the current trial was 0.0667.}
\end{figure}

The distribution of the propensity scores for each Scenario is shown in Figure~\ref{fig:Sim_Box_plot}. 

\subsection{SSR strategy}
Before the initiation of the current study, the required sample size was calculated based on the standard deviation estimated from historical control data. A two-sided significance level of 5\% was used, and the sample size was determined to achieve approximately 80\% power for a presumed true treatment difference $3.5$. In the current study, subjects were randomized in a 2:1 ratio between the treatment and control groups. Sample size re-estimation (SSR) was conducted when 50\% of the initially planned sample size had been accrued. As a supplementary analysis, SSR was also performed at alternative interim points—specifically, when 30\% and 70\% of the planned sample size had been accrued. The corresponding simulation results are presented in Tables~\ref{supp_t:sim_res_typeI} and~\ref{supp_t:sim_res_power} as supplemental results.

\subsection{Evaluation points and hypothesis testing}
To evaluate the proposed method, we considered the approaches based on Strategy 1 and Strategy 2, as well as a baseline scenario in which SSR was not applied. Additionally, to assess the effectiveness of the testing procedure described in Section \ref{sec:test}, we included scenarios in which SSR was applied under Strategy 1 and Strategy 2, but no covariate adjustment was performed thereafter. Specifically, for the methods based on Strategy 1 and Strategy 2 without SSR, hypothesis testing was conducted using the procedure described in Section \ref{sec:test}. In contrast, for the scenarios with SSR followed by no covariate adjustment, a standard two-sample t-test was used.

\subsection{Simulation results}
Simulation results for the Type I error rate and statistical power are presented in Tables~\ref{t:sim_res_typeI} and~\ref{t:sim_res_power}. Regarding the Type I error rate, the proposed methods yielded results comparable to those of the No SSR approach, with Strategy 2 exhibiting better control. In Scenario 5, which involved a substantial discrepancy in covariate distributions, the No SSR method demonstrated relatively better control of the Type I error rate.

Regarding the statistical power, Strategy 1 achieved power nearly equivalent to that of the No SSR approach, whereas Strategy 2 exhibited a modest increase in power. This is attributed to the larger final sample size associated with Strategy 2, because SSR under the strategy 2 tends to yield a greater number of subjects, thereby enhancing the statistical power.

\begin{table}[H]
\centering
\caption{Type I error rates in simulation studies}
\label{t:sim_res_typeI}
\begin{tabular}{|c|c|c|c|c|c|c|c|}
\hline
Strategy 1 & Strategy 2 & No SSR & No ad1 & No ad2 & \# of SS & \# of SSR1 & \# of SSR2\\ \hline
\rowcolor[gray]{0.85}%
\multicolumn{8}{|l|}{Scenario 1: $(\mu_{1h},p_{h},\mu_{3h},\mu_{4h}, \si^2_{1h}, \si^2_{2h}, \si^2_{3h},\si_h^2)=(75,0.5,14,21,8.5^2,2.8^2,3.6^2,10^2)$} \\ \hline 
5.47 & 5.48 & 5.61 & 7.32 & 7.45 & 210.9 & 209.0 & 235.2 \\ \hline
\rowcolor[gray]{0.85}%
\multicolumn{8}{|l|}{Scenario 2: $(\mu_{1h},p_{h},\mu_{3h},\mu_{4h}, \si^2_{1h}, \si^2_{2h}, \si^2_{3h},\si_h^2)=({\bf \textcolor{red}{77}},{\bf \textcolor{red}{0.4}},14,21,8.5^2,2.8^2,3.6^2,10^2)$} \\ \hline 
6.08 & 5.99 & 6.11 & 7.70 & 7.69 & 210.8 & 208.5 & 257.2 \\ \hline
\rowcolor[gray]{0.85}%
\multicolumn{8}{|l|}{Scenario 3: $(\mu_{1h},p_{h},\mu_{3h},\mu_{4h}, \si^2_{1h}, \si^2_{2h}, \si^2_{3h},\si_h^2)=({\bf \textcolor{red}{77}},{\bf \textcolor{red}{0.4}},{\bf \textcolor{red}{15}},{\bf \textcolor{red}{20}},{\bf \textcolor{red}{8.0^2}},{\bf \textcolor{red}{3.2^2}},{\bf \textcolor{red}{3.7^2}},10^2)$} \\ \hline 
6.37 & 6.16 & 6.35 & 7.93 & 8.01 & 215.5 & 210.9 & 268.7 \\ \hline
\rowcolor[gray]{0.85}%
\multicolumn{8}{|l|}{Scenario 4: $(\mu_{1h},p_{h},\mu_{3h},\mu_{4h}, \si^2_{1h}, \si^2_{2h}, \si^2_{3h},\si_h^2)=({\bf \textcolor{red}{73}},{\bf \textcolor{red}{0.6}},{\bf \textcolor{red}{13}},{\bf \textcolor{red}{22}},{\bf \textcolor{red}{8.0^2}},{\bf \textcolor{red}{3.2^2}},{\bf \textcolor{red}{3.7^2}},10^2)$} \\ \hline 
5.48 & 5.35 & 5.43 & 12.14 & 13.49 & 215.4 & 210.8 & 259.1 \\ \hline
\rowcolor[gray]{0.85}%
\multicolumn{8}{|l|}{Scenario 5: $(\mu_{1h},p_{h},\mu_{3h},\mu_{4h}, \si^2_{1h}, \si^2_{2h}, \si^2_{3h},\si_h^2)=({\bf \textcolor{red}{73}},{\bf \textcolor{red}{0.6}},{\bf \textcolor{red}{13}},{\bf \textcolor{red}{22}},{\bf \textcolor{red}{7.0^2}},{\bf \textcolor{red}{2.5^2}},{\bf \textcolor{red}{3.0^2}},{\bf \textcolor{red}{8^2}})$} \\ \hline 
14.03 & 14.45 & 13.27 & 19.92 & 21.48 & 139.9 & 173.7 & 195.7 \\ \hline
\end{tabular} \\
{\footnotesize Strategy 1: Type I error rate based on SSR strategy 1, Strategy 2: Type I error rate based on SSR strategy 2, No SSR: no performing SSR, No ad1: After re-estimating the sample size using SSR strategy 1, a simple t-test was conducted by pooling the data without applying PS weighting, No ad2: After re-estimating the sample size using SSR strategy 2, a simple t-test was conducted by pooling the data without applying PS weighting, \# of SS: The initially calculated sample size before treatment initiation, \# of SSR1: re-estimated sample size based on SSR strategy 1, \# of SSR2: re-estimated sample size based on SSR strategy 2.}
\end{table}

\begin{table}[H]
\centering
\caption{Powers in simulation studies}
\label{t:sim_res_power}
\begin{tabular}{|c|c|c|c|c|c|c|c|}
\hline
Strategy 1 & Strategy 2 & No SSR & No ad1 & No ad2 & \# of SS & \# of SSR1 & \# of SSR2\\ \hline
\rowcolor[gray]{0.85}%
\multicolumn{8}{|l|}{Scenario 1: $(\mu_{1h},p_{h},\mu_{3h},\mu_{4h}, \si^2_{1h}, \si^2_{2h}, \si^2_{3h},\si_h^2)=(75,0.5,14,21,8.5^2,2.8^2,3.6^2,10^2)$} \\ \hline 
87.10 & 90.40 & 87.79 & 90.83 & 93.31 & 210.9 & 213.6 & 235.2 \\ \hline
\rowcolor[gray]{0.85}%
\multicolumn{8}{|l|}{Scenario 2: $(\mu_{1h},p_{h},\mu_{3h},\mu_{4h}, \si^2_{1h}, \si^2_{2h}, \si^2_{3h},\si_h^2)=({\bf \textcolor{red}{77}},{\bf \textcolor{red}{0.4}},14,21,8.5^2,2.8^2,3.6^2,10^2)$} \\ \hline 
85.10 & 90.88 & 85.81 & 91.36 & 95.24 & 210.8 & 213.1 & 256.9 \\ \hline
\rowcolor[gray]{0.85}%
\multicolumn{8}{|l|}{Scenario 3: $(\mu_{1h},p_{h},\mu_{3h},\mu_{4h}, \si^2_{1h}, \si^2_{2h}, \si^2_{3h},\si_h^2)=({\bf \textcolor{red}{77}},{\bf \textcolor{red}{0.4}},{\bf \textcolor{red}{15}},{\bf \textcolor{red}{20}},{\bf \textcolor{red}{8.0^2}},{\bf \textcolor{red}{3.2^2}},{\bf \textcolor{red}{3.7^2}},10^2)$} \\ \hline 
83.35 & 90.27 & 84.48 & 91.59 & 95.88 & 215.4 & 215.3 & 268.6 \\ \hline
\rowcolor[gray]{0.85}%
\multicolumn{8}{|l|}{Scenario 4: $(\mu_{1h},p_{h},\mu_{3h},\mu_{4h}, \si^2_{1h}, \si^2_{2h}, \si^2_{3h},\si_h^2)=({\bf \textcolor{red}{73}},{\bf \textcolor{red}{0.6}},{\bf \textcolor{red}{13}},{\bf \textcolor{red}{22}},{\bf \textcolor{red}{8.0^2}},{\bf \textcolor{red}{3.2^2}},{\bf \textcolor{red}{3.7^2}},10^2)$} \\ \hline 
87.75 & 92.98 & 88.70 & 69.82 & 77.47 & 215.4 & 215.7 & 258.9 \\ \hline
\rowcolor[gray]{0.85}%
\multicolumn{8}{|l|}{Scenario 5: $(\mu_{1h},p_{h},\mu_{3h},\mu_{4h}, \si^2_{1h}, \si^2_{2h}, \si^2_{3h},\si_h^2)=({\bf \textcolor{red}{73}},{\bf \textcolor{red}{0.6}},{\bf \textcolor{red}{13}},{\bf \textcolor{red}{22}},{\bf \textcolor{red}{7.0^2}},{\bf \textcolor{red}{2.5^2}},{\bf \textcolor{red}{3.0^2}},{\bf \textcolor{red}{8^2}})$} \\ \hline 
90.37 & 92.29 & 84.91 & 96.90 & 97.61 & 139.9 & 179.3 & 195.7 \\ \hline
\end{tabular} \\
{\footnotesize Strategy 1: power based on SSR strategy 1, Strategy 2: power based on SSR strategy 2, No SSR: no performing SSR, No ad1: After re-estimating the sample size using SSR strategy 1, a simple t-test was conducted by pooling the data without applying PS weighting, No ad2: After re-estimating the sample size using SSR strategy 2, a simple t-test was conducted by pooling the data without applying PS weighting, \# of SS: The initially calculated sample size before treatment initiation, \# of SSR1: re-estimated sample size based on SSR strategy 1, \# of SSR2: re-estimated sample size based on SSR strategy 2.}
\end{table}

\section{Case study}\label{sec:case}
We demonstrate the re-sample size design using data from two randomized controlled studies: ADC-016 and ADC-027. Both studies were conducted as part of the Alzheimer's Disease Cooperative Study (ADCS). We treat ADC-016 as the historical study and ADC-027 as the current study. A hybrid control design was constructed using ADC-016 as the historical control and ADC-027 as the current study, in which the sample sizes of the docosahexaenoic acid (DHA) and placebo groups were balanced in a 1:1 ratio. The baseline covariates used to estimate the propensity scores were age, sex, education, and mini-mental state examination (MMSE). The outcome of interest was the ADAS score. Assuming a two-sided significance level of 5\%, a mean difference of 3.5, and a standard deviation of 13, a sample size of 220 patients per group is required to achieve statistical power exceeding 80\%. Demographic and baseline characteristics are summarized in Table~\ref{t:summary}, and the analysis of covariance results for ADAS are presented in Table~\ref{t:coeff}.

\begin{table}[H]
\centering
\caption{Summary table}
\label{t:summary}
\begin{tabular}{|c|c|c|c|c|}
\hline
 & historical & \multicolumn{3}{c|}{current}\\ \hline
 & Placebo & Placebo & DHA & Total\\ \hline
N &	169 & 164 & 238 & 402 \\ \hline
Age (yrs) &	77.8 (7.9) & 76.0 (7.8) &	75.9 (9.0) &	75.9 (8.5)\\ \hline
Gender (Male, \%) &	46.2 & 40.2 & 52.9 & 47.8\\ \hline
Education & 13.9 (3.2) & 14.3 (2.7) & 14.3 (2.9) & 14.3 (2.8)\\ \hline
MMSE & 20.9 (3.7) & 20.3 (3.6) & 20.9 (3.6) & 20.7 (3.6)\\ \hline
ADAS & 29.0 (12.5) [9.9] & 33.4 (13.3) [9.8] & 32.7 (13.2) [10.1] & - \\ \hline
\end{tabular} \\
{\footnotesize Mean (SD). [XX.X] refers to the standard deviation of the residuals, presented as a reference value for the variance of the error term in the simulation dataset. Because ADAS is the outcome of interest, the total was not calculated."}
\end{table}

\begin{table}[H]
\centering
\caption{Coefficient parameter}
\label{t:coeff}
\begin{tabular}{|c|c|c|c|}
\hline
 & historical & \multicolumn{2}{c|}{current}\\ \hline
 & Placebo & Placebo & DHA\\ \hline
Intercept &	76.8 & 104.6 & 80.2\\ \hline
Age (yrs) &	$-$0.04 & $-$0.29 &	$-$0.04\\ \hline
Gender (Male, \%) &	1.09 & $-$2.62 & $-$3.17\\ \hline
Education & 0.05 & 0.08 & 0.46\\ \hline
MMSE & $-$2.17 & $-$2.41 & $-$2.33\\ \hline
\end{tabular} \\
\end{table}

Sample size re-estimation was conducted as a case study using data from the ADC-027 study (current study). Patients were ordered according to their enrollment sequence, and three interim time points were considered: when 30\% (Scenario (a)), 50\% (Scenario (b)), and 70\% (Scenario (c)) of the total target enrollment had been accrued. For each scenario, the re-estimation was performed based on the subset of data available at the respective time point. 

\begin{figure}[H]
    \centering
    \includegraphics[width=0.8\linewidth]{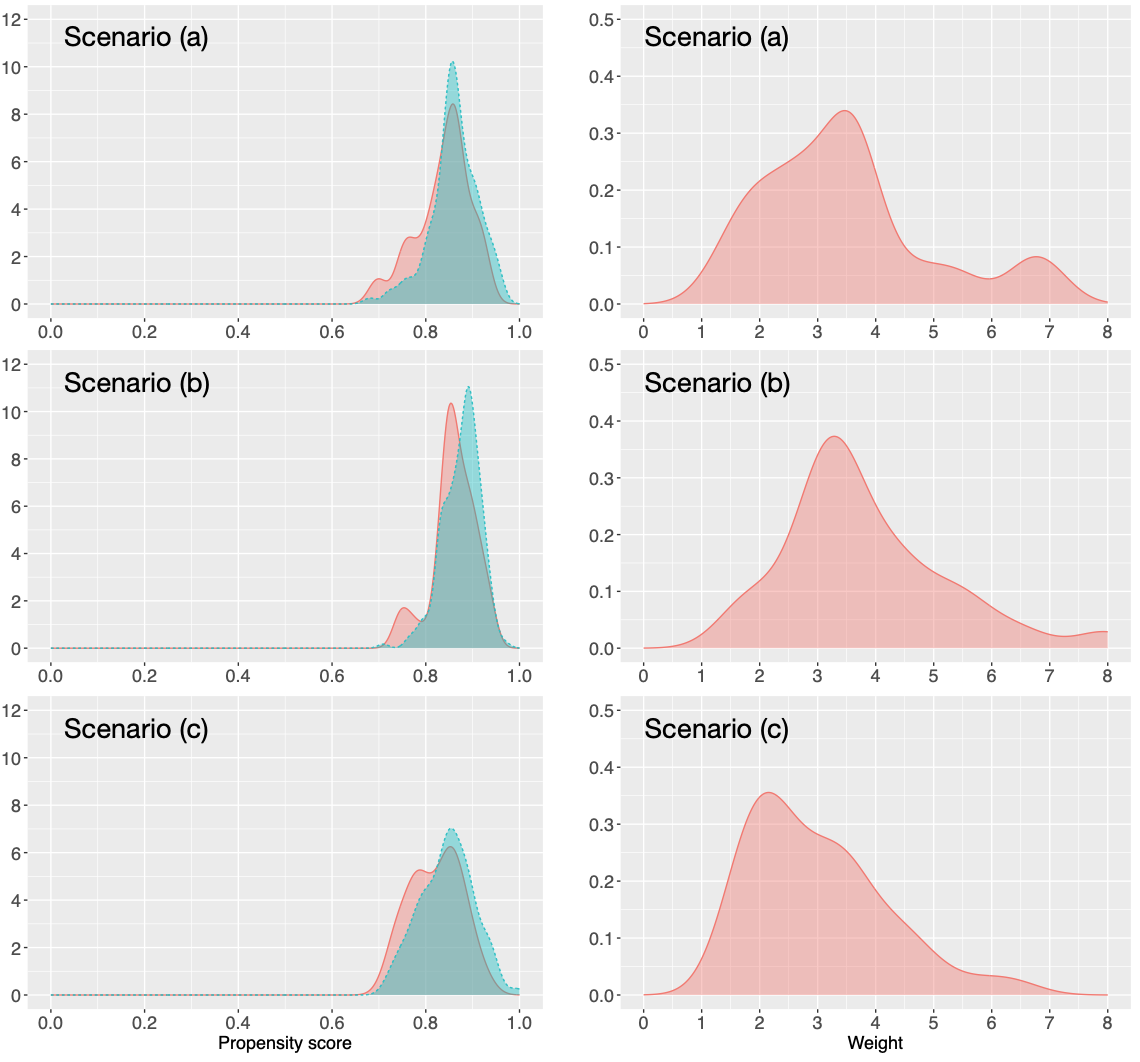}
    \caption{Distributions of propensity scores and weights}
    \label{fig:Box_plot}
    {\footnotesize Green indicates the current trial (dashed line), and red indicates the historical trial. Regarding the weight, the current trial used fixed values, which were not illustrated in the figure. Specifically, the weights in the current trial were 0.583 in scenario (a), 0.572 in scenario (b), and 0.594 in scenario (c).}
\end{figure}

In Scenario (a), 30\% of the patients in the ADC-027 study had been enrolled, comprising 70 patients in the DHA group and 50 patients in the placebo group. A hybrid control design incorporating historical data was employed, and the 20 missing placebo patients in the ADC-027 study were supplemented by randomly sampling from the historical dataset. The distribution of propensity scores for each study is shown in Figure~\ref{fig:Box_plot}. The results of the sample size re-estimation are presented in Table~\ref{t:ssr}.

\begin{table}[H]
\centering
\caption{Sample size re-estimation per one group}
\label{t:ssr}
\begin{tabular}{|c|c|c|c|}\hline
Scenario & strategy 1 & strategy 2 \\ \hline
\multicolumn{3}{|c|}{Initial sample size: 220}\\ \hline
(a) & 227 & 301\\ \hline
(b) & 232 & 258\\ \hline
(c) & 229 & 269\\ \hline
\end{tabular} \\
\end{table}

In Scenario (b), the current study included 115 patients in the DHA group and 86 patients in the placebo group. To compensate for the shortfall of 29 placebo patients, an additional 29 subjects were randomly sampled from the historical data. In Scenario (c), the current study included 167 patients in the DHA group and 114 patients in the placebo group. To compensate for the shortfall of 53 placebo patients, an additional 53 subjects were randomly sampled from the historical data.

For Strategy 1, since the variance changed very little, the required sample size remained nearly unchanged. In contrast, Strategy 2 accounts for differences in the distribution of covariates such as age and sex, which reduces the effective information borrowed from the historical data and results in the enrollment of more patients than initially planned.

\section{Discussion}
We proposed two sample size re-estimation (SSR) strategies for double-blinded randomized studies employing a hybrid control design, both of which preserve blinding of the treatment groups. One of the strategies enables sample size re-estimation without unblinding the primary outcome. Both strategies account for the degree of information borrowing from historical control data by adjusting for differences in the distribution of covariates. The first strategy uses the outcome variance adjusted by IPW, which corrects for differences in baseline characteristics between the current and historical studies. This approach is intended for use by unblinded statisticians, such as members of an independent data monitoring committee, and re-estimates the sample size by replacing the originally assumed variance with the IPW-adjusted variance. The second strategy does not require primary outcome data. Instead, it adjusts the assumed variance for initial sample size determination using only differences in baseline covariates between the current and historical studies, and then re-estimates the sample size accordingly. Because the treatment effect estimation incorporates historical data through IPW, we propose randomly sampling the required number of historical control patients to match the number of control patients needed. The test statistic is constructed based on an estimating equation derived from M-estimation theory.

In the simulation study, we evaluated the Type I error rate and statistical power while varying the distribution of baseline covariates between the historical and current studies. When baseline distributions were aligned across studies for Scenario 1, the Type I error rate was maintained at the nominal level. Even when discrepancies in baseline covariates were present, the proposed SSR methods provided better control of the Type I error rate compared to methods that either did not implement SSR or implemented SSR without IPW adjustment. However, in Scenario 5, when the distribution and variability of covariates were made to differ substantially between the studies, the No SSR approach demonstrated slightly better control of the Type I error rate. A possible explanation is that the extreme differences in covariate distributions between the studies impaired the ability of the propensity score model to adequately distinguish between them, leading to bias. In such cases—where covariate distributions between studies differ markedly—it may be advisable to consider forgoing the borrowing of historical information altogether. A limitation of our approach is that when covariate distributions differed substantially between the current and historical studies, borrowing external control data in a hybrid design led to inflation of the Type I error rate. Therefore, in situations where baseline covariates differ markedly between past and current studies, it may be advisable to avoid borrowing historical information. In the No Adjustment case, where SSR was performed but no IPW adjustment was applied in the hypothesis testing, inflation of the Type I error rate was observed. In Scenario 1, where the covariate distributions were intended to be aligned, exact matching was not achieved because the data were generated based on the distribution of the propensity scores. Therefore, the inflation of the Type I error rate is likely due to the lack of IPW adjustment. With respect to statistical power, Strategy 1 yielded results comparable to the No SSR approach in all scenarios except Scenario 5, as the variability of the outcome did not differ substantially between studies, resulting in the initial sample size being retained even after SSR. In contrast, in Scenario 5, where outcome variability differed between studies, SSR led to an increased sample size and consequently higher power than the No SSR approach. For Strategy 2, higher power was observed across all scenarios compared to No SSR. 

In the case study, we applied the hybrid control design to two real-world randomized controlled studies. Under Strategy 1, which re-estimates sample size based on outcome variance, the re-estimated sample size remained consistent with the original sample size, regardless of whether SSR was performed at 30\%, 50\%, or 70\% enrollment. In contrast, Strategy 2, which adjusts the variance based on differences in baseline characteristics between the historical and current studies, tended to recommend larger sample sizes. This may be attributable to the observed discrepancy in propensity score distributions between the two studies.

The proposed SSR strategies maintained the Type I error rate at the nominal level when baseline covariates were well balanced, and also suppressed Type I error inflation while achieving high statistical power even when the covariates were moderately imbalanced. In particular, Strategy 2—based on IPW and not requiring outcome data—demonstrated high power while maintaining Type I error control comparable to that of a design without SSR. These simulation results support the practical utility of Strategy 2. However, a limitation of Strategy 2 is its consistent tendency to recommend larger sample sizes compared to Strategy 1, due to the absence of outcome information.

In conclusion, we developed a sample size re-estimation framework that accounts for the degree of information borrowing from historical studies at an interim stage of an ongoing study. Simulation results demonstrate that the proposed methods offer improved statistical performance compared to approaches without SSR.


\newpage
\bibliography{main.bib}
\bibliographystyle{unsrt}

\newpage
\begin{appendix}
\section{Derivation of the Test Statistics}
\label{appa}
In this section, we consider the derivation of the test statistic (\ref{Test1}). Specifically, we use the property that the statistic described in Section 2.2 can be regarded as the solution of an M-estimator \cite{St2002}.

We can consider the following estimating equation:
\begin{align}
\label{eq1}
\sum_{i=1}^{n}\left(
\begin{array}{c}
\frac{A_{i}}{{\rm Pr}(A=1)}\left(Y_{i}-\hat{\theta}_{1}\right)\\
\left[\frac{(1-A_{i})}{{\rm Pr}(A=0\mid R=1)}W_{i}^{r=1}+W_{i}^{r=0}\right]\left(Y_{i}-\hat{\theta}_{0}\right)
\end{array}
\right)=\bld{0}.
\end{align}
By conducting the Taylor expansion around the true value $\bld{\theta}^{0}:=\left(\theta^{0}_{1},\theta^{0}_{0}\right)^{\top}$, 
$$
\sqrt{n}\left(\hat{\bld{\theta}}-\bld{\theta}^{0}\right)\approx\frac{1}{\sqrt{n}}\sum_{i=1}^{n}\left(
\begin{array}{c}
\frac{A_{i}}{{\rm Pr}(A=1)}\left(Y_{i}-\theta^{0}_{1}\right)\\
\left[\frac{(1-A_{i})}{{\rm Pr}(A=0\mid R=1)}W_{i}^{r=1}+W_{i}^{r=0}\right]\left(Y_{i}-\theta^{0}_{0}\right)
\end{array}
\right).
$$
Therefore, under some regularity conditions (e.g., Ferguson\cite{Fe2017}),
$$
\sqrt{n}\left(\hat{\bld{\theta}}-\bld{\theta}^{0}\right)\stackrel{d}{\to}N(0,\Sigma),
$$
where
$$
\Sigma=\left(
\begin{array}{cc}
{\rm E}\left[\frac{A}{{\rm Pr}(A=1)^2}\left(Y-\theta^{0}_{1}\right)^2\right]&0\\
0&{\rm E}\left[\left\{\frac{(1-A)}{{\rm Pr}(A=0\mid R=1)}W^{r=1}+W^{r=0}\right\}^2\left(Y-\theta^{0}_{0}\right)^2\right]
\end{array}
\right).
$$
Finally, applying the Delta method\cite{Sh2008}, (\ref{Test1}) can be derived under the Null hypothesis ($\theta_{1}-\theta_{0}=\tau_{0}$).

\newpage
\section{Supplemental simulation results}
\begin{table}[H]
\centering
\caption{Type I error rates in supplemental simulation studies}
\label{supp_t:sim_res_typeI}
\begin{tabular}{|c|c|c|c|c|c|c|c|}\hline
\rowcolor[gray]{0.95}%
\multicolumn{8}{|l|}{Using 30\% of the current study} \\ \hline 
Strategy 1 & Strategy 2 & No SSR & No ad1 & No ad2 & \# of SS & \# of SSR1 & \# of SSR2\\ \hline
\rowcolor[gray]{0.85}%
\multicolumn{8}{|l|}{Scenario 1: $(\mu_{1h},p_{h},\mu_{3h},\mu_{4h}, \si^2_{1h}, \si^2_{2h}, \si^2_{3h},\si_h^2)=(75,0.5,14,21,8.5^2,2.8^2,3.6^2,10^2)$} \\ \hline 
5.59 & 5.52 & 5.72 & 7.29 & 7.44 & 210.9 & 207.5 & 246.2 \\ \hline
\rowcolor[gray]{0.85}%
\multicolumn{8}{|l|}{Scenario 2: $(\mu_{1h},p_{h},\mu_{3h},\mu_{4h}, \si^2_{1h}, \si^2_{2h}, \si^2_{3h},\si_h^2)=({\bf \textcolor{red}{77}},{\bf \textcolor{red}{0.4}},14,21,8.5^2,2.8^2,3.6^2,10^2)$} \\ \hline 
6.20 & 6.16 & 6.12 & 7.53 & 7.89 & 210.9 & 207.2 & 269.4 \\ \hline
\rowcolor[gray]{0.85}%
\multicolumn{8}{|l|}{Scenario 3: $(\mu_{1h},p_{h},\mu_{3h},\mu_{4h}, \si^2_{1h}, \si^2_{2h}, \si^2_{3h},\si_h^2)=({\bf \textcolor{red}{77}},{\bf \textcolor{red}{0.4}},{\bf \textcolor{red}{15}},{\bf \textcolor{red}{20}},{\bf \textcolor{red}{8.0^2}},{\bf \textcolor{red}{3.2^2}},{\bf \textcolor{red}{3.7^2}},10^2)$} \\ \hline 
6.29 & 6.22 & 6.43 & 7.62 & 8.05 & 215.4 & 209.2 & 282 \\ \hline
\rowcolor[gray]{0.85}%
\multicolumn{8}{|l|}{Scenario 4: $(\mu_{1h},p_{h},\mu_{3h},\mu_{4h}, \si^2_{1h}, \si^2_{2h}, \si^2_{3h},\si_h^2)=({\bf \textcolor{red}{73}},{\bf \textcolor{red}{0.6}},{\bf \textcolor{red}{13}},{\bf \textcolor{red}{22}},{\bf \textcolor{red}{8.0^2}},{\bf \textcolor{red}{3.2^2}},{\bf \textcolor{red}{3.7^2}},10^2)$} \\ \hline 
5.43 & 5.45 & 5.45 & 12.26 & 13.88 & 215.4 & 209.3 & 272.5 \\ \hline
\rowcolor[gray]{0.85}%
\multicolumn{8}{|l|}{Scenario 5: $(\mu_{1h},p_{h},\mu_{3h},\mu_{4h}, \si^2_{1h}, \si^2_{2h}, \si^2_{3h},\si_h^2)=({\bf \textcolor{red}{73}},{\bf \textcolor{red}{0.6}},{\bf \textcolor{red}{13}},{\bf \textcolor{red}{22}},{\bf \textcolor{red}{7.0^2}},{\bf \textcolor{red}{2.5^2}},{\bf \textcolor{red}{3.0^2}},{\bf \textcolor{red}{8^2}})$} \\ \hline 
14.24 & 14.65 & 13.56 & 19.82 & 22.03 & 139.9 & 172.4 & 204.9 \\ \hline
\rowcolor[gray]{0.95}%
\multicolumn{8}{|l|}{Using 70\% of the current study} \\ \hline 
\rowcolor[gray]{0.85}%
\multicolumn{8}{|l|}{Scenario 1: $(\mu_{1h},p_{h},\mu_{3h},\mu_{4h}, \si^2_{1h}, \si^2_{2h}, \si^2_{3h},\si_h^2)=(75,0.5,14,21,8.5^2,2.8^2,3.6^2,10^2)$} \\ \hline 
5.66 & 5.48 & 5.59 & 7.25 & 7.53 & 210.8 & 209.1 & 230.5 \\ \hline
\rowcolor[gray]{0.85}%
\multicolumn{8}{|l|}{Scenario 2: $(\mu_{1h},p_{h},\mu_{3h},\mu_{4h}, \si^2_{1h}, \si^2_{2h}, \si^2_{3h},\si_h^2)=({\bf \textcolor{red}{77}},{\bf \textcolor{red}{0.4}},14,21,8.5^2,2.8^2,3.6^2,10^2)$} \\ \hline 
6.03 & 5.94 & 5.97 & 7.50 & 7.85 & 210.8 & 209.2 & 251.8 \\ \hline
\rowcolor[gray]{0.85}%
\multicolumn{8}{|l|}{Scenario 3: $(\mu_{1h},p_{h},\mu_{3h},\mu_{4h}, \si^2_{1h}, \si^2_{2h}, \si^2_{3h},\si_h^2)=({\bf \textcolor{red}{77}},{\bf \textcolor{red}{0.4}},{\bf \textcolor{red}{15}},{\bf \textcolor{red}{20}},{\bf \textcolor{red}{8.0^2}},{\bf \textcolor{red}{3.2^2}},{\bf \textcolor{red}{3.7^2}},10^2)$} \\ \hline 
6.37 & 6.42 & 6.36 & 7.66 & 8.12 & 215.4 & 211.3 & 262.9 \\ \hline
\rowcolor[gray]{0.85}%
\multicolumn{8}{|l|}{Scenario 4: $(\mu_{1h},p_{h},\mu_{3h},\mu_{4h}, \si^2_{1h}, \si^2_{2h}, \si^2_{3h},\si_h^2)=({\bf \textcolor{red}{73}},{\bf \textcolor{red}{0.6}},{\bf \textcolor{red}{13}},{\bf \textcolor{red}{22}},{\bf \textcolor{red}{8.0^2}},{\bf \textcolor{red}{3.2^2}},{\bf \textcolor{red}{3.7^2}},10^2)$} \\ \hline 
5.43 & 5.39 & 5.57 & 12.47 & 13.79 & 215.4 & 211.4 & 253.5 \\ \hline
\rowcolor[gray]{0.85}%
\multicolumn{8}{|l|}{Scenario 5: $(\mu_{1h},p_{h},\mu_{3h},\mu_{4h}, \si^2_{1h}, \si^2_{2h}, \si^2_{3h},\si_h^2)=({\bf \textcolor{red}{73}},{\bf \textcolor{red}{0.6}},{\bf \textcolor{red}{13}},{\bf \textcolor{red}{22}},{\bf \textcolor{red}{7.0^2}},{\bf \textcolor{red}{2.5^2}},{\bf \textcolor{red}{3.0^2}},{\bf \textcolor{red}{8^2}})$} \\ \hline 
13.93 & 14.39 & 13.24 & 19.87 & 21.19 & 139.9 & 174.2 & 191.5 \\ \hline
\end{tabular} \\
{\footnotesize Strategy 1: Type I error rate based on SSR strategy 1, Strategy 2: Type I error rate based on SSR strategy 2, No SSR: no performing SSR, No ad1: After re-estimating the sample size using SSR strategy 1, a simple t-test was conducted by pooling the data without applying PS weighting, No ad2: After re-estimating the sample size using SSR strategy 2, a simple t-test was conducted by pooling the data without applying PS weighting, \# of SS: The initially calculated sample size before treatment initiation, \# of SSR1: re-estimated sample size based on SSR strategy 1, \# of SSR2: re-estimated sample size based on SSR strategy 2.}
\end{table}

\begin{table}[H]
\centering
\caption{Powers in simulation studies}
\label{supp_t:sim_res_power}
\begin{tabular}{|c|c|c|c|c|c|c|c|}\hline
\rowcolor[gray]{0.95}%
\multicolumn{8}{|l|}{Using 30\% of the current study} \\ \hline 
Strategy 1 & Strategy 2 & No SSR & No ad1 & No ad2 & \# of SS & \# of SSR1 & \# of SSR2\\ \hline
\rowcolor[gray]{0.85}%
\multicolumn{8}{|l|}{Scenario 1: $(\mu_{1h},p_{h},\mu_{3h},\mu_{4h}, \si^2_{1h}, \si^2_{2h}, \si^2_{3h},\si_h^2)=(75,0.5,14,21,8.5^2,2.8^2,3.6^2,10^2)$} \\ \hline 
86.49 & 91.32 & 87.84 & 90.30 & 94.14 & 210.9 & 212.2 & 246.4 \\ \hline
\rowcolor[gray]{0.85}%
\multicolumn{8}{|l|}{Scenario 2: $(\mu_{1h},p_{h},\mu_{3h},\mu_{4h}, \si^2_{1h}, \si^2_{2h}, \si^2_{3h},\si_h^2)=({\bf \textcolor{red}{77}},{\bf \textcolor{red}{0.4}},14,21,8.5^2,2.8^2,3.6^2,10^2)$} \\ \hline 
84.73 & 91.62 & 85.61 & 91.02 & 95.70 & 210.9 & 211.7 & 269.4 \\ \hline
\rowcolor[gray]{0.85}%
\multicolumn{8}{|l|}{Scenario 3: $(\mu_{1h},p_{h},\mu_{3h},\mu_{4h}, \si^2_{1h}, \si^2_{2h}, \si^2_{3h},\si_h^2)=({\bf \textcolor{red}{77}},{\bf \textcolor{red}{0.4}},{\bf \textcolor{red}{15}},{\bf \textcolor{red}{20}},{\bf \textcolor{red}{8.0^2}},{\bf \textcolor{red}{3.2^2}},{\bf \textcolor{red}{3.7^2}},10^2)$} \\ \hline 
82.52 & 90.77 & 84.22 & 91.01 & 96.24 & 215.5 & 213.9 & 281.7 \\ \hline
\rowcolor[gray]{0.85}%
\multicolumn{8}{|l|}{Scenario 4: $(\mu_{1h},p_{h},\mu_{3h},\mu_{4h}, \si^2_{1h}, \si^2_{2h}, \si^2_{3h},\si_h^2)=({\bf \textcolor{red}{73}},{\bf \textcolor{red}{0.6}},{\bf \textcolor{red}{13}},{\bf \textcolor{red}{22}},{\bf \textcolor{red}{8.0^2}},{\bf \textcolor{red}{3.2^2}},{\bf \textcolor{red}{3.7^2}},10^2)$} \\ \hline 
87.07 & 93.59 & 88.89 & 69.44 & 79.07 & 215.4 & 213.8 & 272.2 \\ \hline
\rowcolor[gray]{0.85}%
\multicolumn{8}{|l|}{Scenario 5: $(\mu_{1h},p_{h},\mu_{3h},\mu_{4h}, \si^2_{1h}, \si^2_{2h}, \si^2_{3h},\si_h^2)=({\bf \textcolor{red}{73}},{\bf \textcolor{red}{0.6}},{\bf \textcolor{red}{13}},{\bf \textcolor{red}{22}},{\bf \textcolor{red}{7.0^2}},{\bf \textcolor{red}{2.5^2}},{\bf \textcolor{red}{3.0^2}},{\bf \textcolor{red}{8^2}})$} \\ \hline 
89.89 & 92.54 & 84.88 & 96.70 & 97.71 & 139.9 & 178.3 & 204.6 \\ \hline
\rowcolor[gray]{0.95}%
\multicolumn{8}{|l|}{Using 70\% of the current study} \\ \hline 
\rowcolor[gray]{0.85}%
\multicolumn{8}{|l|}{Scenario 1: $(\mu_{1h},p_{h},\mu_{3h},\mu_{4h}, \si^2_{1h}, \si^2_{2h}, \si^2_{3h},\si_h^2)=(75,0.5,14,21,8.5^2,2.8^2,3.6^2,10^2)$} \\ \hline 
87.47 & 89.99 & 87.62 & 91.03 & 93.09 & 210.8 & 213.7 & 230.5 \\ \hline
\rowcolor[gray]{0.85}%
\multicolumn{8}{|l|}{Scenario 2: $(\mu_{1h},p_{h},\mu_{3h},\mu_{4h}, \si^2_{1h}, \si^2_{2h}, \si^2_{3h},\si_h^2)=({\bf \textcolor{red}{77}},{\bf \textcolor{red}{0.4}},14,21,8.5^2,2.8^2,3.6^2,10^2)$} \\ \hline 
85.40 & 90.48 & 85.82 & 91.63 & 95.00 & 210.9 & 213.8 & 251.8 \\ \hline
\rowcolor[gray]{0.85}%
\multicolumn{8}{|l|}{Scenario 3: $(\mu_{1h},p_{h},\mu_{3h},\mu_{4h}, \si^2_{1h}, \si^2_{2h}, \si^2_{3h},\si_h^2)=({\bf \textcolor{red}{77}},{\bf \textcolor{red}{0.4}},{\bf \textcolor{red}{15}},{\bf \textcolor{red}{20}},{\bf \textcolor{red}{8.0^2}},{\bf \textcolor{red}{3.2^2}},{\bf \textcolor{red}{3.7^2}},10^2)$} \\ \hline 
83.73 & 89.94 & 84.20 & 91.81 & 95.71 & 215.4 & 215.8 & 262.8 \\ \hline
\rowcolor[gray]{0.85}%
\multicolumn{8}{|l|}{Scenario 4: $(\mu_{1h},p_{h},\mu_{3h},\mu_{4h}, \si^2_{1h}, \si^2_{2h}, \si^2_{3h},\si_h^2)=({\bf \textcolor{red}{73}},{\bf \textcolor{red}{0.6}},{\bf \textcolor{red}{13}},{\bf \textcolor{red}{22}},{\bf \textcolor{red}{8.0^2}},{\bf \textcolor{red}{3.2^2}},{\bf \textcolor{red}{3.7^2}},10^2)$} \\ \hline 
88.16 & 92.81 & 88.91 & 70.20 & 76.93 & 215.4 & 216.1 & 253.4 \\ \hline
\rowcolor[gray]{0.85}%
\multicolumn{8}{|l|}{Scenario 5: $(\mu_{1h},p_{h},\mu_{3h},\mu_{4h}, \si^2_{1h}, \si^2_{2h}, \si^2_{3h},\si_h^2)=({\bf \textcolor{red}{73}},{\bf \textcolor{red}{0.6}},{\bf \textcolor{red}{13}},{\bf \textcolor{red}{22}},{\bf \textcolor{red}{7.0^2}},{\bf \textcolor{red}{2.5^2}},{\bf \textcolor{red}{3.0^2}},{\bf \textcolor{red}{8^2}})$} \\ \hline 
90.68 & 92.13 & 84.72 & 97.15 & 97.48 & 139.9 & 179.9 & 191.4 \\ \hline
\end{tabular} \\
{\footnotesize Strategy 1: power based on SSR strategy 1, Strategy 2: power based on SSR strategy 2, No SSR: no performing SSR, No ad1: After re-estimating the sample size using SSR strategy 1, a simple t-test was conducted by pooling the data without applying PS weighting, No ad2: After re-estimating the sample size using SSR strategy 2, a simple t-test was conducted by pooling the data without applying PS weighting, \# of SS: The initially calculated sample size before treatment initiation, \# of SSR1: re-estimated sample size based on SSR strategy 1, \# of SSR2: re-estimated sample size based on SSR strategy 2.}
\end{table}

\end{appendix}

\end{document}